\documentclass[11pt,twoside]{article}
\usepackage{asp2010}

\resetcounters

\begin{document}

\aspvoltitle{Four Decades of Research on Massive Stars}
\aspcpryear{2012}
\aspvolauthor{Carmelle Robert, Nicole St-Louis and Laurent Drissen, eds.}

\title{An X-ray Survey of Colliding Wind Binaries} 
\author{Marc~Gagn\'e$^1$, Garrett~Fehon$^1$, Michael R. Savoy$^1$, Carlos A. Cartagena$^1$, David H. Cohen$^2$, and Stanley P. Owocki$^3$ 
\affil{$^1$Dept.\,of Geology \& Astronomy, West Chester Univ., West Chester, PA, USA}
\affil{$^2$Dept.\,of Physics \& Astronomy, Swarthmore College, Swarthmore, PA, USA}
\affil{$^3$Bartol Research Institute, University of Delaware, Newark, DE, USA}
}

\begin{abstract}
We have compiled a list of 36 O+O and 89 Wolf-Rayet binary candidates in the Milky Way and Magellanic clouds detected
with the {\it Chandra}, {\it XMM-Newton} and {\it ROSAT} satellites to probe the connection 
between their X-ray properties 
and their system characteristics.
Of the WR binaries with published parameters,
all but two have $kT>0.9$~keV. 
The most X-ray luminous WR binaries are typically very long period systems.
The WR binaries show a nearly four-order of magnitude spread in X-ray luminosity, even among
among systems with very similar WR primaries.
Among the O+O binaries, short-period systems 
generally have soft X-ray spectra 
and longer period systems show harder X-ray spectra 
again with a large spread in $L_{\rm X}/L_{\rm bol}$.
\end{abstract}

\section{Introduction}
X-rays from non-magnetic massive stars are thought to be produced two ways: via embedded wind shocks in the radiately driven wind close to the star,
and, in massive binaries, via shocks in the wind collision zone between the two stars \citep{1992ApJ...386..265S}.
3D numerical simulations of colliding wind shocks in $\eta$~Carinae and the WC8 binary WR 140 \citep{2010AAS...21542602R, 2011ApJ...726..105P}
correctly predict the characteristic rise, rapid decline, and recovery of the X-ray light curve as these highly eccentric, long-period, adiabatic
systems approach and emerge from periastron. 3D simulations of O+O binaries by \citep{2010MNRAS.403.1657P} reproduce the overall X-ray luminosity and post-shock
temperatures of a number of systems spanning a range of mass-loss rates, orbital periods and eccentricities.
In particular, they were able to produce the strong, but relatively soft X-ray emission seen in some highly radiative, short-period systems. 
On the other hand, the 3D model for WR~22 (WN7h + O9 III-V) over-predicts
the observed $L_{\rm X}$ by an order of magnitude or more \citep{2011A&A...530A.119P}. 

The O+O binaries in the {\it Chandra} Carina Complex project showed a wide range of $L_{\rm X}/L_{\rm bol}$ \citep{2011ApJS..194....7N}, and
\citep{2011ApJS..194....5G} found that the short-period systems have significantly softer X-ray spectra than the longer-period systems. 
As Owocki discusses in these proceedings, thin-shell mixing may play in important role in setting the scaling between $L_{\rm X}$ and $L_{\rm bol}$
in the winds of single stars, and could produce significant cooling in the wind collision zone of close, massive binaries.
Motivated by these results, we undertook a survey of all known WR and O+O binaries with X-ray fluxes measured with
{\it Chandra} or {\it XMM-Newton}.

\section{Methodology}
To begin, we searched the literature for {\it Chandra} and {\it XMM} analyses of WR and O+O binaries. To those, we added
X-ray sources in the XMM-Newton Serendipitous Source Catalog (2XMMi), in the {\it XMM-Newton} XAssist Source List, or in the {\it Chandra} XAssist Source List,
within $15^{\prime\prime}$ of positions in the 7th catalog of WR stars \citep{2001yCat.3215....0V}
and O+O binaries in the SB9 catalog of spectroscopic binaries \citep{2009yCat....102020P}.
For the O+O binaries with reliable X-ray fluxes, column densities, and
distances, we calculated a 0.5--8 keV X-ray flux. These results and the corresponding X-ray, optical, and distance references are reported in Table 1.
Reference codes are noted in parentheses in the references section.
Similarly, results for the known WR binaries are reported in Table 2.
In some cases, WR binary X-ray luminosities were taken from the {\it ROSAT} survey of \citet{2000MNRAS.318..214I}.
In cases where stars (e.g., WR~101k) in the {\it Chandra} or {\it XMM} XAssist source lists were detected in many observations,
or by multiple cameras on {\it XMM}, we report $L_{\rm X}$ based on a median unabsorbed X-ray flux.

\begin{table} 
\caption{Known X-ray Luminous O+O binaries}
\smallskip
\begin{center}
{\scriptsize
\begin{tabular}{lllcccccl}
\tableline
\noalign{\smallskip}
Name & Primary & Second. & Dist. & Period & $kT$ & $L_{\rm X}$ & $L_{\rm X}/L_{\rm bol}$ & Ref. \\
     & Sp.Type  & Sp.Type  & (kpc) & (days) & (keV) & (cgs) &         &  Note   \\
\noalign{\smallskip}
\tableline
\noalign{\smallskip}
   HD 215835     &   O5.5 V((f))         &   O6.5 V((f))         & 3.5   & 2.11  & 0.3   & 33.73         &   $-$6.16     &          aly  \\
      Mk33Na     &        O6.5 V         &        O3 If*         & 51    & 1140  & 1.3   & 33.72         &   $-$6.08     &            l  \\
  Cyg OB2 9      &         O5 If         &          O6-7         & 1.2   & 852.9         & 1.2   & 33.52         &   $-$6.30     &          dmz  \\
    HD 47129     &        O7.5 I         &          O6 I         & 1.5   & 14.4  & 1.3   & 33.36         &   $-$6.02     &            kq  \\
      CEN 1B     &            O4         &             ?         & 1.6   &      $\ldots$         & 2.3   & 33.28         &   $-$6.00     &            it  \\
   HD 165052     &        O6.5 V         &        O6.5 V         & 1.6   & 2.95  & 0.6   & 33.18         &   $-$6.00     &         alwyp\AA  \\
    HD 93250     &    O4 III(fc)         &             O         & 2.3   & 250   & 2.3   & 33.18         &   $-$6.41     &           su1  \\
      CEN 1A     &            O4         &             ?         & 1.6   &      $\ldots$         & 6.5   & 33.16         &   $-$6.12     &            it  \\
   HD 101436     &        O6.5 V         &          O7 V         & 2.3   & 37.37         &      $\ldots$         & 33.11         &   $-$6.17     &            b  \\
    HD 93403     &        O5.5 I         &          O7 V         & 2.3   & 15.093        & 1     & 33.11         &   $-$6.41     &          sxu  \\
   HD 159176     &          O6 V         &          O6 V         & 0.8   & 3.36677       & 0.3   & 33.07         &   $-$5.89     &           ael  \\
   HD 101131     &   O6.5 V((f))         &        O8.5 V         & 2.3   & 9.65  &      $\ldots$         & 32.98         &   $-$6.33     &            b  \\
    V729 Cyg     &          O7 f         &          O6 f         & 1.8   & 6.5978        & 0.6   & 32.96         &   $-$6.74     &            p  \\
   HD 101205     &  O7 IIIn((f))         &             O         & 2.3   & 2.45  &      $\ldots$         & 32.95         &   $-$6.54     &            b  \\
     HD 1337     &        O9 III         &        O9 III         & 3.9   & 3.52  &      $\ldots$         & 32.90  &   $-$6.65     &           al  \\
 Cyg OB2 8A      &     O5.5 I(f)         &            O6         & 1.2   & 21.907        & 1     & 32.88         &   $-$6.75     &           mnz  \\
    MT91 516     &        O5.5 V         &             ?         & 1.2   &      $\ldots$         & 0.5   & 32.73         &   $-$6.79     &            mz  \\
   HD 101413     &          O8 V         &         B3: V         & 2.3   & 150   &      $\ldots$         & 32.68         &   $-$6.09     &            b  \\
   HD 100213     &          O7 V         &          O8 V         & 2.1   & 1.38729       &      $\ldots$         & 32.63         &   $-$5.95     &            al  \\
    HD 93205     &          O3 V         &          O8 V         & 2.3   & 6.0803        & 0.3   & 32.55         &   $-$6.82     &         \$sux  \\
      QZ Car     &   O9.7 Ib:(n)         &        O9 III         & 2.3   & 20.72         & 1     & 32.55         &   $-$7.26     &           sN  \\
   HD 101190     &    O4 V((f+))         &      O7-7.5 V         & 2.3   & 6.05  &      $\ldots$         & 32.52         &   $-$6.74     &            b  \\
 CPR2002 A11     &      O7.5 Ibf         &             ?         & 1.2   &      $\ldots$         & 1.6   & 32.47         &   $-$6.64     &            mz  \\
    HD 57060     &          O7 f         &            O7         & 1.5   & 4.39  & 0.7   & 32.43         &   $-$7.38     &          alx  \\
    HD 97484     &          O7.5         &          O8.5         & 3.2   & 3.41428       & 0.6   & 32.41         &   $-$6.87     &             p  \\
   HD 206267     &   O6.5 V((f))         &            O9         & 0.8   & 3.71  & 0.6   & 32.36         &   $-$6.89     &           alp  \\
   HD 152218     &         O9 IV         &        O9.7 V         & 1.4   & 5.604         & 0.6   & 32.05         &   $-$6.73     &            pv  \\
   HD 93161A     &          O8 V         &          O9 V         & 2.3   & 8.566         & 0.5   & 31.94         &   $-$6.92     &          osu  \\
   HD 152218     &         O9 IV         &        O9.7 V         & 1.6   & 5.6   & 0.5   & 31.93         &   $-6.85$     &           pv  \\
   Wd 1 30      &    O9-B0.5 Ia         &             ?         & 4.0   & $\ldots$ & 1.3 & 31.90 & $\ldots$ & \O$\hbar$ \\
   Tr 16-110     &          O7 V         &     O8 V, O9 V        & 2.3   & 3.62864       & 0.6   & 31.74         &   $-$7.24     &          sxu  \\
    HD 93343     &          O8 V         &      O7-8.5 V         & 2.3   & 44.15         & 3.2   & 31.66         &   $-$6.98     &          sxu  \\
    Tr 16-34     &          O8 V         &        O9.5 V         & 2.3   & 2.9995        & 0.6   & 31.56         &   $-$7.23     &          gsu  \\
   Tr 16-104     &          O7 V         &   O9.5 V, B0.2        & 2.3   & 2.1529        & 0.5   & 31.38         &   $-$7.29     &         jsxu  \\
       FO 15     &        O5.5 V         &        O9.5 V         & 2.3   & 1.41356       & 0.5   & 31.24         &   $-$7.65     &          rsu  \\
     Tr 16-1     &        O9.5 V         &        B0.3 V         & 2.3   & 1.4693        & 0.3   & 30.87         &   $-$7.30     &          sxu  \\
\noalign{\smallskip}
\tableline
\end{tabular}
}
\end{center}
\end{table}

\begin{table} 
\caption{Known X-ray Luminous WR binaries}
\smallskip
\begin{center}
{\scriptsize
\begin{tabular}{lllcccccl}
\tableline
\noalign{\smallskip}
Name & Primary & Second. & Dist. & Period & $kT$ & $L_{\rm X}$ & $L_{\rm X}/L_{\rm bol}$ & Ref. \\
     & Sp.Type  & Sp.Type & (kpc) & (days) & (keV) & (cgs) &  &  \\
\noalign{\smallskip}
\tableline
\noalign{\smallskip}
      WR 48a  &        WC8ed  &            ?  &      3.8  &  7800  &     2.3  &    35.39  & $-$4.00    &             @  \\
        Mk34  &       WN6(h)  &            ?  &     51  &  1134  &     2.3  &    35.38  & $-$4.72  &              3l  \\
  $\eta$ Car  &          LBV  &            O  &      2.3  &  2024  &     4.4  &    35.26  & $-$5.02  &               7  \\
      R 140a  &          WC5  &          WN4  &     51  &   880       &     0.9  &    35.25  & $-$4.65  &          l  \\
       WR 25  &         WN6h  &         O4 f  &      2.3  &   207.8       &     1.3  &    35.11  & $-$5.49  &        3M  \\
      R 136c  &         WN5h  &            ?  &     51  &   998  &     3.0    &    35.04  & $-$4.96  &           3lQ  \\
CXO J1745-28  &         WN9h  &           O?  &      7.6  &   189  &     2.7  &    35.04  & $-$4.78  &               3WP  \\
       WR 28  &       WN6(h)  &          OB?  &     10.8  &     $\ldots$  &   $\ldots$ &    34.86  &   $\ldots$ &     F  \\
      WR 43c  &      WN6+abs  &            ?  &      7.6  &     8.89      &   $\ldots$ &    34.85  & $-$5.33  &      KV3  \\
      WR 140  &        WC7pd  &         O4-5  &      1.1  &  2900  &   $\ldots$ &    34.68  & $-$4.75  &            GHJS  \\
      Mk33Sa  &          WC5  &     O3 IIIf*  &     51  &  1120  &     0.6  &    34.63  & $-$4.87  &                 l  \\
     Brey 16  &         WN4b  &          O5:  &     51  &    18  &     7.0    &    34.58  & $-$4.30  &               BT  \\
      WR 43a  &        WN6ha  &            ?  &     10.1  &     3.772     &   $\ldots$ &    34.30  & $-$6.30  &        I  \\
      R 136a  &          WN5  &               &     51  &     $\ldots$  &     1.8  &    34.28  & $-$5.92  &        l  \\
       WR 29  &         WN7h  &            O  &     17.2  &     3.16415   &   $\ldots$ &    34.21  &   $\ldots$ &      FH4  \\
     WR 101k  &       WN9-11  &            ?  &      8.0  &     9.72      &   $\ldots$ &    34.12  &   $\ldots$ &        J  \\
        Mk39  &          WN6  &        O3 If  &     51  &    92.6       &     1.6  &    34.11  & $-$5.89  &       3l  \\
     HD 5980  &          WN3  &           OB  &     61  &    19.266     &     7.0    &    34.08  & $-$5.90  &        T  \\
     WR 121a  &          WN7  &         a/OB  &      5.6  &     $\ldots$  &   $\ldots$ &    34.07  &   $\ldots$ &        J  \\
   Arches-F6  &         WN9h  &            ?  &      8.0  &     $\ldots$  &     1.9  &    34.04  &   $\ldots$ &       RU  \\
     R 136a3  &         WN5h  &               &     51  &     $\ldots$  &     4.2  &    33.93  &   $\ldots$ &        3  \\
      WR 20a  &        WN6ha  &        WN6ha  &      8.0  &     3.68      &     0.5  &    33.90  & $-$6.15  &        X  \\
       WR 65  &         WC9d  &           OB  &      5.0  &     $\ldots$  &   $\ldots$ &    33.90  &   $\ldots$ &        J  \\
   Arches-F7  &         WN9h  &            ?  &      8.0  &     $\ldots$  &     2.1  &    33.86  &   $\ldots$ &       RU  \\
      WR 20b  &        WN6ha  &            ?  &      8.0  &     $\ldots$  &     3.6  &    33.81  & $-$6.16  &       3X  \\
    Brey 10a  &    O3 If*/WN6  &               &     51  &     3.23      &   $\ldots$ &    33.80  &   $\ldots$ &        3  \\
       WR 87  &         WN7h  &           OB  &      2.9  &     $\ldots$  &   $\ldots$ &    33.75  & $-$5.75  &       FG  \\
       WR 93  &          WC7  &         O7-9  &      2.5  &     $\ldots$  &   $\ldots$ &    33.74  &   $\ldots$ &        J  \\
    BAT99-32  &       WN6(h)  &               &     51  &     1.91      &   $\ldots$ &    33.70  &   $\ldots$ &        3  \\
   Arches-F9  &         WN9h  &               &      8.0  &     $\ldots$  &     3.3  &    33.66  & $-$6.22  &          RU  \\
     Brey 26  &       WN6(h)  &            ?  &     51  &     1.91      &   $\ldots$ &    33.65  &   $\ldots$ &        0  \\
       WR 71  &          WN6  &          OB?  &      9.0  &     7.69      &   $\ldots$ &    33.63  &   $\ldots$ &       AF  \\
       WR 63  &          WN7  &           OB  &      3.9  &     $\ldots$  &   $\ldots$ &    33.63  &   $\ldots$ &        F  \\
       R 144  &          WN6  &            ?  &     51  &     $\ldots$  &   $\ldots$ &    33.52  &   $\ldots$ &      30Q  \\
     Av 336a  &           WN  &           O6  &     61  &    19.56      &     2.2  &    33.52  &   $\ldots$ &        T  \\
       WR 35  &         WN6h  &          OB?  &     17.9  &     $\ldots$  &   $\ldots$ &    33.47  & $-$5.51  &        F  \\
      WR 145  &      WN7/WCE  &            ?  &      1.2  &    20         &   $\ldots$ &    33.43  &   $\ldots$ &        J  \\
       WR 51  &          WN4  &          OB?  &      8.1  &     $\ldots$  &   $\ldots$ &    33.26  & $-$5.82  &        F  \\
     Brey 56  &          WN6  &            ?  &     51  &     $\ldots$  &     2.3  &    33.23  & $-$5.75  &       BT  \\
       WR 66  &       WN8(h)  &          cc?  &      3.3  &     3.515     &   $\ldots$ &    33.21  & $-$6.17  &        F  \\
       WR 48  &          WC6  &   O9.5/B0 Iab  &      2.2  &    18.341     &   $\ldots$ &    33.20  &   $\ldots$ &        H  \\
       R 134  &       WN6(h)  &            ?  &     51  &   786  &     1.1  &    33.18  & $-$6.92  &                    l  \\
       WR 11  &          WC8  &   O7.5 III-V  &      0.3  &    78.53      &     1.0    &    33.17  & $-$5.39  &      GHJ  \\
        Mk42  &          WN6  &        O3 If  &     51  &   922  &     1.1  &    33.15  & $-$6.85  &                l  \\
        Mk30  &          WN6  &       O3 If*  &     51  &     4.7       &   $\ldots$ &    33.11  &   $\ldots$ &        0  \\
       R 139  &           WN  &       O6 Iaf  &     51  &   952  &     1.8  &    33.08  & $-$6.92  &                 l  \\
   Wd 1 WR B  &         WN7o  &            ?  &    4.0  &  3.52 & 1.4     & 33.05 &   $\ldots$ & \O$\hbar$ \\
       WR 47  &          WN6  &         O5 V  &      3.8  &     6.2393    &     1.1  &    33.05  & $-$6.55  &         AHp2  \\
       WR 67  &          WN6  &          OB?  &      3.3  &     $\ldots$  &   $\ldots$ &    33.04  & $-$5.64  &        F  \\
       R 145  &         WN6h  &            ?  &     51  &   158.8       &     1.6  &    33.00    &   $\ldots$ &       Q6  \\
      WR 133  &          WN5  &         O9 I  &      2.1  &   112.4       &   $\ldots$ &    33.00    & $-$6.36  &      EHp  \\
      WR 21a  &          WN6  &          O/a  &      3.0  &    31.673     &     3.3  &    33.00    & $-$5.78  &      OY3  \\
       WR 22  &         WN7h  &     O9 III-V  &      2.3  &    80.336     &     1.4  &    32.95  & $-$6.90  &     3Hp5  \\
      WR 158  &         WN7h  &          Be?  &      7.9  &     $\ldots$  &   $\ldots$ &    32.93  & $-$6.55  &        F  \\
      R 140b  &         WN6h  &            ?  &     51  &     2.76      &   $\ldots$ &    32.90  &   $\ldots$ &        T  \\
       WR 89  &         WN8h  &           OB  &      2.9  &     $\ldots$  &   $\ldots$ &    32.90  & $-$6.98  &       F  \\
      WR 46   &         WN3p  &          OB?  &      4.1  &     0.2825    &   $\ldots$ &    32.87  & $-$6.22  &       CZ \\
       R 135  &         WN7h  &            ?  &     51  &     2.11      &   $\ldots$ &    32.78  &   $\ldots$ &        0  \\
      WR 148  &         WN8h  &     B3 IV/BH  &      8.3  &     4.317364  &   $\ldots$ &    32.78  & $-$6.80  &       FH  \\
      WR 132  &          WC6  &            ?  &      3.9  &     8.16      &   $\ldots$ &    32.75  & $-$5.93  &        F  \\
       WR 36  &        WN5-6  &          OB?  &      8.5  &     $\ldots$  &   $\ldots$ &    32.74  & $-$6.14  &        F  \\
      WR 146  &          WC6  &           O8  &      1.2  &  1235  &   $\ldots$ &    32.73  &   $\ldots$ &            J \\
       WR 24  &        WN6ha  &            ?  &      2.3  &     $\ldots$  &     1.7  &    32.71  & $-$6.93  &        8  \\
     Brey 65  &        WN7ha  &            ?  &     51  &     3  &   $\ldots$ &    32.70  &   $\ldots$ &          30  \\
     \noalign{\smallskip}
\tableline
\end{tabular}
}
\end{center}
\end{table}

\setcounter{table}{1}
\begin{table} 
\caption{Known X-ray Luminous WR binaries (continued)}
\smallskip
\begin{center}
{\scriptsize
\begin{tabular}{lllcccccl}
\tableline
\noalign{\smallskip}
Name & Primary & Second. & Dist. & Period & $kT$ & $L_{\rm X}$ & $L_{\rm X}/L_{\rm bol}$ & Ref. \\
     & SpType  & SpType & (kpc) & (days) & (keV) & (cgs) &  &  \\
\noalign{\smallskip}
\tableline
\noalign{\smallskip}
     WR 147N  &       WN8(h)  &       B0.5 V  &      0.6  &  2880  &     1.8  &    32.67  & $-$7.01  &                p9  \\
       WR 44  &          WN4  &          OB?  &     10.0  &     $\ldots$  &   $\ldots$ &    32.64  & $-$6.54  &        F  \\
      WR 155  &          WN6  &     O9 II-Ib  &      2.8  &     1.641244  &   $\ldots$ &    32.64  & $-$6.44  &      EHp  \\
      WR 108  &         WN9h  &           OB  &      5.6  &     $\ldots$  &   $\ldots$ &    32.62  & $-$6.76  &        F  \\
       WR 1  &          WN4  &            ?  &      0.7  &     6.1       &   $\ldots$ &    32.50  &   $\ldots$ &       X!  \\
     WR 139   &          WN5  &     O6 III-V  &      1.9  &     4.212435  &     3.3  &    32.50  & $-$6.40  &       EHZ2  \\
      WR 125  &        WC7ed  &       O9 III  &      3.1  &  6600       &   $\ldots$ &    32.49  & $-$6.29  &         F  \\
       WR 12  &         WN8h  &            ?  &      5.0  &    23.923     &   $\ldots$ &    32.48  & $-$6.90  &      3AF  \\
      WR 114  &          WC5  &          OB?  &      2.0  &     $\ldots$  &   $\ldots$ &    32.41  & $-$5.87  &        F  \\
      WR 138  &          WN5  &           B?  &      1.3  &  1538  &   $\ldots$ &    32.20  & $-$6.68  &                GH  \\
       WR 79  &          WC7  &         O5-8  &      2.0  &     8.8908    &   $\ldots$ &    32.18  &   $\ldots$ &        HJ  \\
        WR 6  &         WN4b  &            ?  &      0.9  &     3.765     &   $\ldots$ &    32.12  &   $\ldots$ &        H  \\
      WR 115  &          WN6  &          OB?  &      2.0  &     $\ldots$  &   $\ldots$ &    32.08  & $-$6.90  &       FG  \\
      WR 141  &          WN5  &     O5 V-III  &      1.3  &    21.6895    &   $\ldots$ &    31.99  & $-$7.98  &      EHX  \\
       WR 39  &          WC7  &          OB?  &      5.5  &     $\ldots$  &   $\ldots$ &    31.92  & $-$6.76  &        F  \\
        WR 3  &          WN3  &           O4  &      5.9  &    46.85      &   $\ldots$ &    31.91  & $-$7.27  &        F \\
       WR 14  &          WC7  &            ?  &      2.0  &     2.42      &   $\ldots$ &    31.90  & $-$6.58  &        F  \\
      WR 128  &       WN4(h)  &          OB?  &      9.4  &     3.56      &   $\ldots$ &    31.75  & $-$7.33  &        F  \\
       WR 86  &          WC7  &     B0 III-I  &      2.9  &     $\ldots$  &   $\ldots$ &    31.72  & $-$7.36  &        F \\
      WR 136  &         WN6h  &            ?  &      1.6  &     $\ldots$  &     2.2  &    31.51  & $-$7.47  &        8  \\
      WR 121  &         WC9d  &            ?  &      1.8  &     $\ldots$  &   $\ldots$ &    31.49  & $-$7.40  &       EL  \\
        WR 4  &          WC5  &            ?  &      2.4  &     2.4096    &   $\ldots$ &    31.37  & $-$7.21  &       FH  \\
      WR 143  &          WC4  &          OB?  &      1.1  &     $\ldots$  &   $\ldots$ &    31.22  & $-$7.36  &        F \\
   Wd 1 WR F  &          WC9  &         OB+?  &      4.0  &  5.05   & 18     & 31.14 &   $\ldots$ & $\hbar$\ss \\
   Wd 1 WR L  &        WN9h:  &            ?  &      4.0  &  $<10$   & 8     & 30.88 &   $\ldots$ & $\hbar$ \\
\noalign{\smallskip}
\tableline
\end{tabular}
}
\end{center}
\end{table}

\section{Results and Discussion}

We emphasize that the results presented in Tables 1 and 2 are preliminary. Moreover, mass-loss rates, orbital parameters,
and accurate X-ray spectral parameters are needed for a number of systems. Nonetheless, it is clear
that the most X-ray luminous WR binaries, like the LBV binary $\eta$~Car,
are typically very long period systems. The exceptions, which include the 8.9-day WN6 binary WR~43c=NGC~3603-A1
with $\log L_{\rm X}/L_{\rm bol} = -5.3$, are remarkable, and merit further study.
We note that only two WR systems have $kT < 0.9$~keV: Mk33Sa (WC5 + O3 IIIf*) in the LMC and WR 20a (WN6ha + WN6ha) in Westerlund~2.
Though the spectral type of the secondary is often not known, we note that 
WR systems with known early-O and supergiant secondaries often have $\log L_{\rm X} > 33$.
Other systems, e.g., WR~101k, which was observed repeatedly as part of the {\it XMM} and {\it Chandra} galactic center surveys, are variable
from observation to observation. WR 48a, the most X-ray luminous WR binary in Table 2, has undergone a dramatic decline in
X-ray flux in 2011 in the {\it Swift} XRT (A. M. T. Pollock, private communication).

Because of their lower mass-loss rates, the O+O binaries in Table 1 have far lower $L_{\rm X}$, with $\log L_{\rm X}/L_{\rm bol}$
in the range $-5.9$ to $-7.7$ than the WR stars in Table 2. Short-period O+O systems ($P < 10$~days) have soft X-ray spectra ($kT < 0.8$~keV) and
longer period systems show harder X-ray spectra ($kT > 1$~keV). This suggests that in close O+O binaries,
the higher density shocks, on average, undergo significant cooling, e.g., as a result of thin-shell mixing.
For O+O systems with $\log L_{\rm X}/L_{\rm bol} < 7$, embedded wind shocks may account for a large fraction
of the X-ray luminosity.

\nocite{*}
\bibliography{mgagne-v3}

\end{document}